\documentstyle[aps,prl,multicol]{revtex}
\include{epsf}
\begin{document}
\preprint{\today}
\draft
\title{Neutron scattering in molecular liquids: \\ Influence of 
orientational degrees of freedom and the prepeak in a fragile glass
former} 

\author{Christoph Theis and Rolf Schilling}
\address{Institut f\"ur Physik, Johannes Gutenberg--Universit\"at,
Staudinger Weg 7, D--55099 Mainz, Germany}
\maketitle

\begin{abstract}
The intermediate scattering function $S_{n}(q,t)$ for 
neutron-scattering is expanded with respect to a complete set of
correlation functions which describe the dynamical correlations in a
molecular liquid.
For the static ns--structure factor $S_{n}(q)$ of a system of
diatomic molecules the results of the expansion are compared with the
exact results from a MD--simulation and it is shown that the 
Sears--expansion, which is 
commonly used to
interpret such data, fails in the supercooled regime. 
The representation for $S_{n}(q)$ is used to draw
conclusions about the $q$--dependence and especially the origin of the
prepeak.
\end{abstract}

\pacs{PACS numbers: 61.12.-q, 61.20.-p, 61.25.Em}

\begin{multicols}{2}
Neutron scattering is one of the most important tools to determine
the structure and dynamics of condensed matter.
For instance, it has been applied widely in the study of supercooled
liquids and the glass--transition, (see e.g. ref. \cite{richter89}
and for more recent work \cite{toelle97,wuttke95}),
where it has been of great value to test theories such as the
mode--coupling--theory (see e.g. \cite{goetze91,schilling94}).
One of the major advantages of neutron scattering is that the neutron 
as an electrically
neutral probe is not influenced by the electron cloud of the target
atoms but interacts only with the nucleus. Consequently neutron 
scattering can
be directly interpreted in terms of the atomic motion. However, if one
wants to examine a molecular system not only with respect to the
motion of the constituent atoms but in terms of the molecular units
some care has to be taken. Since the superposition of scattering from
atomic sites is in general not equivalent to scattering from the
molecular center of mass, neutron scattering from molecular liquids is
sensitive to orientational as well as translational correlations in
the system.
Attempts to account for this fact in the analysis of ns--data include
the expansion into "partial waves" proposed by Sears \cite{sears67}.
Originally conceived for liquids with negligible correlations between
rotational and translational motion it was shown recently to describe
well the incoherent scattering in supercooled water
\cite{chen97}. The expansion proposed in the present Letter is closely
related to that of Sears but does not make any assumptions on the
strength of the coupling between rotation and translation. 
As a consequence it is
able to describe well also the coherent scattering in the 
{\em supercooled}
regime. In addition it offers interesting new aspects for the
interpretation of the q--dependence of the static
ns--structurefactor $S_{n}(q)$.

A feature of special interest in supercooled liquids and glasses is
the appearance of a prepeak in the static ns-structure factor at a
$q$--value that corresponds to distances larger than the average
nearest--neighbour distance. As a sign of intermediate range order
prepeaks have been studied in a variety of systems (see e.g. 
\cite{wilson94,misawa90,iyetomi93,boerjesson93}). Mostly they have been
attributed to the network structure of strong liquids but as 
our
present analysis of a system of rigid diatomic molecules shows 
prepeaks can
appear even in the most simple molecular systems.

We consider a set of N rigid molecules of identical geometry each
consisting of s atoms. 
The assumption of rigidity is one that is commonly used as well in the
theoretical analysis as in computer simulations of the liquid state.
It is justified on the ground that at the temperature of the liquid
only the lowest vibrational states are populated.
The starting point of our analysis is the
following site--site representation of the intermediate scattering 
function
(cf. \cite{sears67})
\begin{eqnarray}
\label{sneutron}
& &S_{n}(q,t) = 
\frac{1}{Ns} \sum_{j,j'=1}^N \sum_{\nu,\nu'=1}^s \; \times \\ 
& & \quad \times \, \left(
a_{coh}^{j \nu} \, a_{coh}^{j' \nu'} + a_{inc}^{j \nu
\, 2} \delta_{j j'} \, \delta_{\nu \nu'} \right) 
\, \left\langle e^{i
{\bf q} \cdot ({\bf R}_{j \nu}(t) - {\bf R}_{j' \nu'})} \right\rangle
\nonumber
\end{eqnarray}
where $a_{coh}^{j \nu}$ ($a_{inc}^{j \nu}$) is the coherent 
(incoherent) scattering length and ${\bf R}_{j \nu}(t)$ the position
at time $t$ of atom $\nu$ in molecule $j$. The brackets $\langle ..
\rangle$ denote the canonical average over initial conditions.
Unlike the geometry which
is identical for all molecules the scattering lengths are allowed to
differ from molecule to molecule. This assumption is quite realistic
since the chemical structure is independent of the isotopic
composition whereas the scattering lengths are. The translational and
orientational motion are separated by introducing center of
mass ${\bf R}_j(t)$ 
and relative coordinates ${\bf r}_{j \nu}(t)$, 
i.e. ${\bf R}_{j \nu}(t) = {\bf R}_j
(t)+{\bf r}_{j \nu}(t)$.
Analogously to Sears \cite{sears67} we use
the Rayleigh-expansion of the plane waves 
\begin{eqnarray}
\label{rayleigh}
& & e^{i {\bf q} \cdot {\bf r}_{j \nu}(t)} =  
\sum_{l n} \left[ 4 \pi
(2l+1) \right]^\frac{1}{2} \, i^l \, \times \\
& & \quad \times \, j_l(q r_\nu) \,
Y_l^n(\theta_\nu,\phi_\nu) \, {\cal D}_{0 n}^{l \ast}(\Omega_j(t))
\nonumber
\end{eqnarray}
\noindent
to express the relative coordinates ${\bf r}_{j \nu}(t)$ by the polar
coordinates ($r_\nu,\theta_\nu,\phi_\nu$) of atom $\nu$ with respect
to a body fixed frame with origin at the center of mass of
molecule $j$ 
and by the Euler--angles $\Omega_j(t)$ denoting the
orientation of the body--fixed system with respect to the laboratory~
frame. 
\end{multicols}
\begin{twocolumn}
\noindent
Without loss of generality the $z$--axis of the laboratory
system has been chosen to point in direction of~${\bf q}$.
\noindent
The special functions appearing in eq.(\ref{rayleigh}) are the
spherical Bessel functions $j_l$, the spherical harmonics $Y_l^m$ and
the Wigner functions ${\cal D}_{m n}^l$. Conventions are chosen
according to the textbook of Gray and Gubbins \cite{gray84}. The range
of indices is $l,l'=0,1,2,...$; $-l \le m,n \le l$; $-l' \le m,n' \le
l'$. 
Thus we establish a connection between the intermediate scattering
function $S_{n}(q,t)$ of neutron scattering and the set of functions
\begin{eqnarray}
\label{Sjlm}
& & S_{l n, l' n'}^{m}(q,t) = i^{l'-l} \left[ (2l+1)(2l'+1) 
\right]^{\frac{1}{2}} \frac{1}{N} \sum_{j,j'} \, \times \\
& & \quad \times \, \left\langle e^{-i q (R_j^z(t) - R_{j'}^z)} \, 
{\cal D}_{m
n}^l(\Omega_j(t)) \, {\cal D}_{m n'}^{l' \ast}(\Omega_{j'})
\right\rangle
\nonumber
\end{eqnarray}
which can be split up into self ($j=j'$) and distinct part
($j\not=j'$) giving
\begin{equation}
\label{Sjlmsplit}
S_{l n, l' n'}^{m}(q,t) = S_{l n, l' n'}^{m (s)}(q,t) + S_{l n, l'
n'}^{m (d)}(q,t) .
\end{equation}

This set of correlation functions is the generalization to
arbitrary molecules of the correlators recently introduced for a
single linear molecule in an isotropic liquid \cite{franosch97} and
for molecular liquids of linear molecules
\cite{schilling97}. Similar correlation functions have been used
previously in the study of molecular liquids \cite{gray84,hansen76}.
The
tensors (\ref{Sjlm}) have a number of symmetry
properties which are discussed in refs. \cite{theis97,schilling97}.
We just want
to point out that $S_{0 0,0 0}^0(q,t)$ is the usual intermediate 
scattering
function for the center of mass whereas $S_{l n, l' n'}^m(q,t)$ 
for $l$ or
$l'$ different from one reflect correlations between higher
"multipoles" of the microscopic density.

Inserting eq.(\ref{rayleigh}) into (\ref{sneutron}) one obtains with 
(\ref{Sjlm}) 
the final result:
\begin{eqnarray}
\label{result}
& & S_{n}(q,t) = 
\sum_{l l'} \sum_{n n'} \left[ b_{l n, l'
n'}^{inc}(q) \, S_{l n, l' n'}^{0 (s)}(q,t) \, + \right. \\ 
& & \quad + \, b_{l n, l'n'}^{coh}(q) \, S_{l n, l' n'}^{0 (d)}
(q,t) \left. \right] \nonumber
\end{eqnarray}
with the coefficients
\begin{eqnarray}
\label{binc}
& & b_{l n, l'n'}^{inc}(q) =
\frac{1}{s} \sum_{\nu,\nu'} 4\pi \, j_l(q r_\nu) \, 
j_{l'}(r_{\nu'}) \, \times \\
& & \quad Y_l^{n \ast}(\theta_\nu,\phi_\nu) \, Y_{l'}^{n'}
(\theta_{\nu'},\phi_{\nu'}) \, 
\left[ \overline{a_{coh}^\nu 
a_{coh}^{\nu'}} + \overline{a_{\mbox{inc}}^{\nu \, 2}} \,
\delta_{\nu \nu'} \right] \nonumber \\
\label{bcoh}
& & b_{l n, l'n'}^{coh}(q) = 
\frac{1}{s} \sum_{\nu,\nu'} 4\pi \, j_l(q r_\nu) \, 
j_{l'}(r_{\nu'}) \, \times \\
& & \quad Y_l^{n \ast}(\theta_\nu,\phi_\nu) \, Y_{l'}^{n'}
(\theta_{\nu'},\phi_{\nu'}) \, 
\overline{a_{coh}^{\nu^{}}} \; \overline{a_{coh}^{\nu'}} . 
\nonumber
\end{eqnarray}
Here $\overline{x} = 1/N \sum_j x^j$ denotes
the average over molecules.
A similar result for diluted linear molecules in an isotropic liquid
is given in ref.\cite{franosch97}.
In contrast to the original approach by
Sears we will {\em not} invoke factorization of correlations between
(i) all translational and
rotational motion and (ii) the orientational degrees of freedom of
different molecules. This asumption which is quite common in the
analysis of neutron as well as light--scattering turns out to be not
quite satisfactory in the supercooled regime as we will demonstrate in
the following.
The expansion (\ref{result}) has a number of interesting properties.
Immediately obvious is that eq.(\ref{result}) can be
Fourier--transformed to give a corresponding relation between the
spectra or susceptibilities
Further, we notice that only correlators
with $m=0$ enter into $S_{n}(q,t)$ (which is a consequence of the
{\em isotropy} of the fluid) and that the
coefficients $b_{l n, l'n'}(q)$ are completely determined by the
geometry of the molecule via ($r_\nu,\theta_\nu,\phi_\nu$) and
averages of the scattering lenght of the constituent atoms. The
dependence on the wavevector $q$ enters through the spherical Bessel
functions $j_l(q r_\nu)$ and is thus connected to the radii $r_\nu$
and the value of $l$. This connection between $q$ and $l$ will be very
useful to justify the cutoff of the summations over $l$ in eq.
(\ref{result}) and to analyse the ns--structure factor.
\begin{figure}[htb]
\epsfxsize=75mm
\epsfysize=40mm
\centerline{\hspace{0.75cm} \epsffile{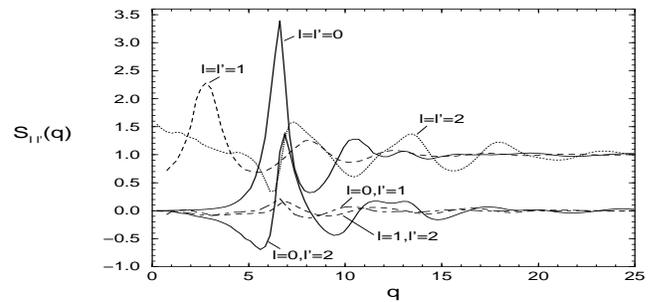}}
\caption{\label{input}
Static molecular correlation functions up to
$l,l' \le 2$ for the lowest temperature $T=0.48$ (in units of the
Lennard--Jones energy of the A--atoms) of the simulation} 
\end{figure}
To demonstrate the quality of expansion (\ref{result}) we consider the
special case of diatomic rigid molecules. From a recent MD--simulation
\cite{kaemmerer97} we have taken the static molecular correlation
functions according to eq.(\ref{Sjlm}) and we have evaluated the
exact result for the static ns--structurefactor using 
eq.(\ref{sneutron}). In that simulation a system of 500 molecules was
considered each consisting of two atoms labeled A and B. The
molecular interaction was represented as the superposition of
Lennard--Jones potentials between the atoms. The molecular bond
length was
fixed at $d=0.5$ in units of the Lennard--Jones radius of the
A--atoms which we will use throughout this Letter. Further details
about the simulation can be found in references \cite{kaemmerer97}.
Since for linear molecules the third Euler--angle $\chi$ doesn't play
a role the distinct parts of the molecular correlation functions 
are nonzero for $n=n'=0$
only and the summations over $n,n'$ can be carried out for the self
part.
Therefore we will skip the indices $n,n'$ as well as the
superscript $m=0$ in the following. The static correlation functions
$S_{l l'}(q)$ evaluated up to $l,l' \le l_{co} = 2$ are shown in
Figure \ref{input} for the lowest temperature of the simulation.
For $l,l' \ge l_{co}$ only the self part $S_{l l'}^{(s)}(q) =
\delta_{l l'}$ is considered in the following calculations.
Specializing to linear diatomic molecules and the static case
eqn.(\ref{result}-\ref{bcoh}) simplify to:
\begin{eqnarray}
\label{snsresult}
& & S_{n}(q) \cong \sum_{l, l' \le l_{co}} b_{l l'}^{coh}(q) \, S_{l
l'}^{(d)}(q) + b^{inc}(q) \\
\label{cohff}
& & b_{l l'}^{coh}(q) = \left[ (2l+1)(2l'+1) \right]^{\frac{1}{2}} \,
j_l(\frac{qd}{2}) \, j_{l'}(\frac{qd}{2}) \; \times \\
& & \quad \times \; \frac{1}{2} \left(
\overline{a_{coh}^{A}} + (-1)^l \overline{a_{coh}^{B}} \right)
\left( \overline{a_{coh}^{A}} + (-1)^{l'} \overline{a_{coh}^{B}}
\right) \nonumber \\
\label{incff}
& & b^{inc}(q) = \frac{1}{2} \left( \overline{a_{coh}^{A 2}} +
\overline{a_{inc}^{A 2}} + \overline{a_{coh}^{B 2}} +
\overline{a_{inc}^{B 2}} \right) +  \\
& & \quad \overline{a_{coh}^{A}} \;
\overline{a_{coh}^{B}} \, j_0(qd) \nonumber 
\end{eqnarray}
The reader should notice that $l_{co}=0$ just yields the usual 
Sears--expansion with the molecular
structure factor $b^{inc}(q)$ and the intermolecular form factor $b_{0
0}^{coh}(q)$ for uncorrelated rotation. In Figure \ref{comparison}
we show the comparison between the exact $S_{n}(q)$ and the
expansion (\ref{snsresult}) for $l_{co}=0$ (Sears), $l_{co}=1$ and
$2$ at the lowest temperature of the simulation. 
The expansion for $l_{co}=2$ is in {\em perfect}
agreement with the exact $S_{n}(q)$ up to $q \approx 9$ and in
qualitative agreement up to almost $q \approx 18$ whereas the
Sears--expansion shows hardly more than qualitative agreement up to
$q \approx 8$. If the scattering length of A-- and B--atoms are
equal it follows from (\ref{cohff}) that only even $l$ and $l'$
contribute to $S_{n}(q)$. Hence $S_{n}(q)$ from (\ref{snsresult}) with
$l_{co}=1$ is identical to the Sears--result. 
If neutron--scattering can
distinguish between A-- and B--atoms the $l_{co}=1$--result is still
not much better than the Sears--expansion (cf. Fig.
(\ref{comparison})) except for the prepeak at $q \approx 3$
which does not show up in the latter (see inset of Fig. 
(\ref{comparison})). This failure of the
Sears--expansion clearly shows that in the supercooled regime the
assumption of uncorrelated rotation of different molecules does not
hold. Upon increasing the temperature these correlations become
weaker and we find that the contributions coming 
from $S_{l l'}(q)$ with
$l,l'$ different from 0 get less important.
Whereas usual ways to account for orientational correlations are
restricted to assuming specific orientational models like in the
analysis of Dore et al. \cite{dore75} or to fitting the data under the
assumption of a preferred orientation \cite{misawa89} our approach is
an exact expansion (provided $l_{co}=\infty$) that makes no ad hoc 
assumptions about
orientational correlations. Of course the calculation of $S_{l n,l'
n'}^m(q)$ can not be done exactly, but requires either approximation
schemes \cite{gray84,hansen76} or simulational methods. In the case
of the time--dependent correlators $S_{l n,l' n'}^m(q,t)$ one may use
the results following from the molecular mode coupling theory
\cite{schilling97}.

Comparing the q--dependence of the input--data in Figure
\ref{input} with that of $S_{n}(q)$ reveals the origin of the
different peaks in the ns--data. Whereas the position of the prepeak
coincides with the first peak in $S_{1 1}(q)$ the main peak has its
origin in the center of mass correlations $S_{0 0}(q)$ though it is
slightly shifted to higher $q$ due 
to the contributions from $S_{0 2}(q)$
and $S_{2 2}(q)$. The peaks at $q \approx 13$ and $q \approx 18$ are
easily recognized as stemming mainly from the corresponding ones in the
correlator $S_{2 2}(q)$. 
\begin{figure}[htb]
\epsfxsize=80mm
\epsfysize=40mm
\centerline{\epsffile{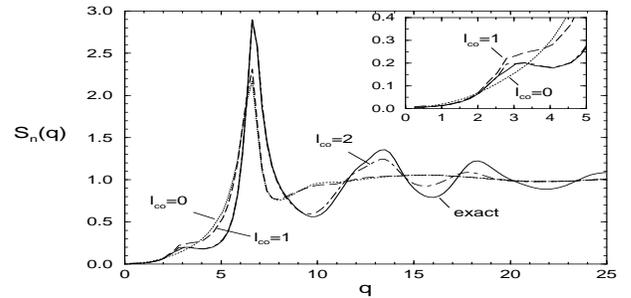}}
\caption{\label{comparison}
Comparison of the different approximations to $S_{n}(q)$ at $T=0.48$ 
for  
$a_{coh}^A=1.4$, $a_{coh}^B=0.25$, $a_{inc}^A=a_{inc}^B=0$.
The inset shows an enlargment of the prepeak region.} 
\end{figure}
\noindent
These obvious identifications are supported even more 
by a closer inspection of the coefficients $b_{l
l'}^{coh}(q)$ which will also explain why the structure in $S_{n}(q)$
at $q \ge 18$ is missed by the cut-off we have chosen. As stated in the
discussion of the general result (\ref{result}) the $q$--dependence
of $b_{l l'}(q)$ is closely related to the values of $l$ and $l'$
through the Bessel functions $j_l(\frac{qd}{2})$. These are
oscillating functions that decay to zero for increasing argument
\cite{abramowitz65}. Whereas $j_0(x)$ has a maximum at $x=0$ the
functions $j_l(x)$ with $l \ge 1$ start at zero and the position of
the first maximum shifts to higher x with increasing l. This property
is inherited by the coefficients $b_{l l'}^{coh}(q)$ and leads to the
consequence that correlators with small $l$ give a contribution at
low $q$ whereas big $l$ determine the behaviour at high $q$ (apart
from the incoherent contribution from $b^{inc}(q)$). 
This is also the
reason why the expansion with $l_{co}=2$ works less good for $q \ge 9$
and particularly for $q \ge
18$ since higher values of $l$ and $l'$ would be needed. 
A quantitative analysis of the
coefficients at the position $q \approx 3$ of the prepeak shows that
about 80\% of the prepeak amplitude originates from $S_{1 1}(q)$.
Therefore the appearance of the prepeak will depend sensitively on the
ratio of the scattering lengths, 
as demonstrated in Figure~\ref{prepeak}.

In the final part of this Letter we want to analyse further the
appearance of a prepeak in $S_{n}(q)$ for diatomic LJ--molecules. As
we have already pointed out in the discussion of the $q$--dependence
of $S_{n}(q)$ the prepeak is connected with the main maximum of the
correlator $S_{1 1}(q)$. The center of mass correlations show no
structure at all at $q \approx 3$ which is a remarkable contrast to
the situation encountered in supercooled water where the intermediate
range order shows up as a prepeak in $S_{0 0}(q)$ \cite{sciortino97}.
This indicates that different mechanisms are responsible for its
existence in the present fragile glassformer
and the network--former. Taking into account the
shape of the spherical harmonic function $Y_1^0 \propto D_{0 0}^1$ one
can conclude that $S_{1 1}(q)$ can be viewed as the correlations
between molecules with parallel orientation 
(of next nearest neighbours). Thus the connection of
intermediate range order with $S_{1 1}(q)$ is in good agreement with
the findings of Misawa \cite{misawa89} who proposed a staggered
parallel orientation as the preferred orientation of nearest neighbours
in liquid halogenes. In another paper Misawa \cite{misawa90} showed 
that the  existence of a preferred orientation can give rise to a
prepeak depending on the ratio of interatomic spacing and bond length.
\begin{figure}[htb]
\epsfxsize=80mm
\epsfysize=40mm
\centerline{\epsffile{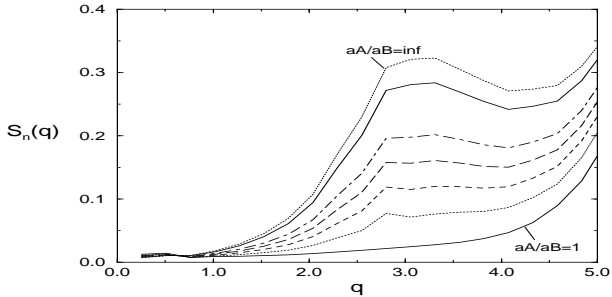}}
\caption{\label{prepeak}Dependence of the prepeak amplitude on the 
ratio of coherent
scattering lengths of A-- and B--atoms.}
\end{figure}
In conclusion we can say that the proposed expansion for the
intermediate scattering function of neutron--scattering offers a
general way of describing orientational correlation effects. It
indicates interesting connections between the $q$--dependence of
$S_{n}(q)$, the orientational correlations for different $l$
and the molecular geometry which are useful in the
interpretation of neutron-scattering results for molecular liquids.
In particular, we have shown that orientational correlations play a
crucial role in the supercooled regime. This is consistent with the
recent results by Bermejo et al. \cite{bermejo97}. These correlations
are also responsible for the existence of a prepeak, whose intensity
depends both on the scattering lengths and the temperature.

\vspace{0.5cm}
\noindent
{\bf Acknowledgement:}
We thank E.Bartsch for pointing out some of the
references given below and we are greatful for the financial support
by SFB--262.

\end{twocolumn}


\vspace{1cm}
\begin{references}
\bibitem{richter89} {\em Dynamics of Disordered Materials} eds.
D.Richter, A.J.Dianoux, W.Petry and J.Teixeira, (Springer, Berlin,
1989).
\bibitem{toelle97} A.T\"olle, H.Schober, J.Wuttke, F.Fujara, 
Phys.Rev. E {\bf 56}, 809, (1997).
\bibitem{wuttke95} J.Wuttke, W.Petry, C.Coddens, F.Fujara,
Phys.Rev. E {\bf 52}, 4026, (1995).
\bibitem{goetze91} W.G\"otze in {\em Liquids, Freezing and the
Glass Transition}, Eds. J.-P.Hansen, D.Levesque and
J.Zinn-Justin (North--Holland, Amsterdam, 1991), p. 287. 
\bibitem{schilling94} R.Schilling in {\em Disorder Effects on 
Relaxational Processes}, Eds. R.Richert, A.Blumen (Springer, 
Berlin, 1994), p.194.
\bibitem{sears67} V.F.Sears, Can.J.Phys. {\bf 45}, 237, (1967).
\bibitem{chen97} S.-H.Chen, P.Gallo, F.Sciortino, P.Tartaglia, 
Phys.Rev. E {\bf 56}, 4231, (1997).
\bibitem{wilson94} M.Wilson and P.A.Madden, Phys.Rev.Lett. {\bf
72}, 3033, (1994).
\bibitem{misawa90} M.Misawa, J.Chem.Phys. {\bf 93}, 6774, (1990).
\bibitem{iyetomi93} H.Iyetomi, P.Vashishta, Phys.Rev. B {\bf 47},
3063, (1993).
\bibitem{boerjesson93} L.B\"orjesson, A.K.Hassan, 
J.Swenson, L.M.Torell, 
A.Fontana, Phys.Rev.Lett. {\bf 70}, 1275, (1993). 
\bibitem{gray84} C.G.Gray und K.E.Gubbins, {\em Theory of molecular
fluids, Volume 1}, (Clarendon Press, Oxford) (1984).
\bibitem{franosch97} T.Franosch, M.Fuchs, W.G\"otze, M.R.Mayr, 
A.P.Singh, Phys.Rev. E {\bf 56}, 5659, (1997).
\bibitem{schilling97} R.Schilling and T.Scheidsteger, Phys.Rev.
E {\bf 56}, 2932, (1997).
\bibitem{hansen76} J.P.Hansen and I.R.Mcdonald, {\em Theory of 
Simple Liquids}, (Academic, London, 1976).
\bibitem{theis97} Diplomarbeit C.Theis, 
Johannes Gutenberg--Universit\"at Mainz, (1997), (unpublished).
\bibitem{kaemmerer97} S.K\"ammerer, W.Kob and R.Schilling, 
Phys.Rev. E {\bf 56}, 5450, (1997), \\
S.K\"ammerer, W.Kob and R.Schilling, Phys.Rev. E 
{\bf 58}, in press, (1998),\\
S.K\"ammerer, W.Kob and R.Schilling, Phys.Rev. E
{\bf 58}, in press, (1998).
\bibitem{dore75} J.C.Dore, G.Walford, D.I.Page, Mol.Phys. {\bf 29}, 565,
(1975).
\bibitem{misawa89} M.Misawa, J.Chem.Phys. {\bf 91}, 2575, (1989).
\bibitem{abramowitz65} M.Abramowitz and I.A.Stegun, {\em Handbook of
Mathematical Functions}, (Dover, New York, 1972).
\bibitem{sciortino97} F.Sciortino, L.Fabbian, S.-H.Chen, P.Tartaglia,
Phys.Rev. E {\bf 56}, 5397, (1997).
\bibitem{bermejo97} F.J.Bermejo et al., Phys.Rev.B {\bf 56}, 11536,
(1997).
\end{references}
\end{document}